\documentstyle[12pt,epsfig]{article}

\catcode`\@=11
\long\def\@makefntext#1{
\protect\noindent \hbox to 3.2pt {\hskip-.9pt
$^{{\ninerm\@thefnmark}}$\hfil}#1\hfill}		

\def\@makefnmark{\hbox to 0pt{$^{\@thefnmark}$\hss}}  

\def\ps@myheadings{\let\@mkboth\@gobbletwo
\def\@oddhead{\hbox{}
\rightmark\hfil\ninerm\thepage}
\def\@oddfoot{}\def\@evenhead{\ninerm\thepage\hfil
\leftmark\hbox{}}\def\@evenfoot{}
\def\sectionmark##1{}\def\subsectionmark##1{}}

\renewcommand{\thefootnote}{\fnsymbol{footnote}}

\newcounter{sectionc}\newcounter{subsectionc}\newcounter{subsubsectionc}
\renewcommand{\section}[1] {\vspace*{0.6cm}\addtocounter{sectionc}{1}
\setcounter{subsectionc}{0}\setcounter{subsubsectionc}{0}\noindent
	{\normalsize\bf\thesectionc. #1}\par\vspace*{0.4cm}}
\renewcommand{\subsection}[1] {\vspace*{0.6cm}\addtocounter{subsectionc}{1}
	\setcounter{subsubsectionc}{0}\noindent
	{\normalsize\it\thesectionc.\thesubsectionc. #1}\par\vspace*{0.4cm}}
\renewcommand{\subsubsection}[1]
{\vspace*{0.6cm}\addtocounter{subsubsectionc}{1}
	\noindent {\normalsize\rm\thesectionc.\thesubsectionc.\thesubsubsectionc.
	#1}\par\vspace*{0.4cm}}

\newcounter{appendixc}
\newcounter{subappendixc}[appendixc]
\newcounter{subsubappendixc}[subappendixc]

\renewcommand{\appendix}[1] {\vspace*{0.6cm}
        \refstepcounter{appendixc}
        \setcounter{figure}{0}
        \setcounter{table}{0}
        \setcounter{equation}{0}
        \renewcommand{\thefigure}{\Alph{appendixc}.\arabic{figure}}
        \renewcommand{\thetable}{\Alph{appendixc}.\arabic{table}}
        \renewcommand{\theappendixc}{\Alph{appendixc}}
        \renewcommand{\theequation}{\Alph{appendixc}.\arabic{equation}}
        \noindent{\bf Appendix \theappendixc #1}\par\vspace*{0.4cm}}

\def\abstracts#1{{

\centering{\begin{minipage}{12.2truecm}\footnotesize\baselineskip=12pt\noindent
	\centerline{\footnotesize ABSTRACT}\vspace*{0.3cm}
	\parindent=0pt #1
	\end{minipage}}\par}}

\renewenvironment{thebibliography}[1]
	{\begin{list}{\arabic{enumi}.}
	{\usecounter{enumi}\setlength{\parsep}{0pt}
\setlength{\leftmargin 1.25cm}{\rightmargin 0pt}
	 \setlength{\itemsep}{0pt} \settowidth
	{\labelwidth}{#1.}\sloppy}}{\end{list}}

\newcounter{itemlistc}
\newcounter{romanlistc}
\newcounter{alphlistc}
\newcounter{arabiclistc}

\newcommand{\fcaption}[1]{
        \refstepcounter{figure}
        \setbox\@tempboxa = \hbox{\footnotesize Fig.~\thefigure. #1}
        \ifdim \wd\@tempboxa > 6in
           {\begin{center}
        \parbox{6in}{\footnotesize\baselineskip=12pt Fig.~\thefigure. #1}
            \end{center}}
        \else
             {\begin{center}
             {\footnotesize Fig.~\thefigure. #1}
              \end{center}}
        \fi}

\newcommand{\tcaption}[1]{
        \refstepcounter{table}
        \setbox\@tempboxa = \hbox{\footnotesize Table~\thetable. #1}
        \ifdim \wd\@tempboxa > 6in
           {\begin{center}
        \parbox{6in}{\footnotesize\baselineskip=12pt Table~\thetable. #1}
            \end{center}}
        \else
             {\begin{center}
             {\footnotesize Table~\thetable. #1}
              \end{center}}
        \fi}

\def\@citex[#1]#2{\if@filesw\immediate\write\@auxout
	{\string\citation{#2}}\fi
\def\@citea{}\@cite{\@for\@citeb:=#2\do
	{\@citea\def\@citea{,}\@ifundefined
	{b@\@citeb}{{\bf ?}\@warning
	{Citation `\@citeb' on page \thepage \space undefined}}
	{\csname b@\@citeb\endcsname}}}{#1}}

\newif\if@cghi
\def\cite{\@cghitrue\@ifnextchar [{\@tempswatrue
	\@citex}{\@tempswafalse\@citex[]}}
\def\citelow{\@cghifalse\@ifnextchar [{\@tempswatrue
	\@citex}{\@tempswafalse\@citex[]}}
\def\@cite#1#2{{$\null^{#1}$\if@tempswa\typeout
	{IJCGA warning: optional citation argument
	ignored: `#2'} \fi}}

 1
 1
 1

\font\ninerm=cmr9

\begin{document}
\begin{flushright}
{\footnotesize
ADP-95-37/T191
}
\end{flushright}

\centerline{\normalsize\bf CHARGE SYMMETRY VIOLATION}
\baselineskip=22pt
\centerline{\normalsize\bf IN NUCLEAR PHYSICS}
\baselineskip=16pt

\centerline{\footnotesize A. W. THOMAS\footnote{ Invited talk presented
at the International Conference on Weak and Electromagnetic Interactions
in Nuclei, Osaka, June 12-16, 1995.}}
\baselineskip=13pt
\centerline{\footnotesize\it Department of Physics and Mathematical
Physics, University of Adelaide,}
\baselineskip=12pt
\centerline{\footnotesize\it Adelaide, S.A. 5005, Australia}
\centerline{\footnotesize E-mail: athomas@physics.adelaide.edu.au}
\vspace*{0.3cm}
\centerline{\footnotesize and}
\vspace*{0.3cm}
\centerline{\footnotesize K. SAITO}
\baselineskip=13pt
\centerline{\footnotesize\it Physics Division, Tohoku College of
Pharmacy}
\baselineskip=12pt
\centerline{\footnotesize\it Sendai 981, Japan}
\centerline{\footnotesize E-mail: ksaito@nucl.phys.tohoku.ac.jp}

\vspace*{0.9cm}
\abstracts{ The study of charge symmetry violation  in nuclear physics
is a potentially enormous subject. Through a few topical examples we aim
to show that it is not a subject of peripheral interest but rather goes
to the heart of our understanding of hadronic systems.}

\normalsize\baselineskip=15pt
\setcounter{footnote}{0}
\renewcommand{\thefootnote}{\alph{footnote}}
\section{Introduction}
The concept of charge symmetry (CS) is not as familiar as that of charge
independence. Whereas the latter requires that the Hamiltonian, $H$,
commutes with all the generators of rotations in isospace ($[H,I_i] =
0, \forall i=1,2,3$), the former requires only that $H$ be invariant
under rotations by $180^o$ about the 2-axis in isospace:
\begin{equation}
[H, e^{i \pi I_2}] = 0 .
\end{equation}
Thus while isospin is frequently broken at the level of a few percent,
CS is often good to a fraction of a percent\cite{mns}. For example, the mass
splitting between the proton ($p$) and the neutron ($n$) is only a
$0.1\%$ effect.

The classic place to test CS is in nucleon-nucleon (NN) scattering. This
is an area where there have recently been some very important new
experiments and some fairly significant new theoretical ideas. Section 2
is devoted to these issues. In section 3 we review recent developments
in the treatment of charge symmetry violation (CSV) in mirror nuclei --
the Okamato-Nolen-Schiffer anomaly. We specifically discuss recent
quark-based treatments of this effect. In this meeting quite a lot of
attention was devoted to the use of nuclear data to extract the element
$V_{ud}$ of the Cabibbo-Kobayashi-Maskawa matrix, in order to test
whether the matrix is unitary. In section 4 we outline some recent work
which suggests the apparent violation of unitarity may be (at least)
partially explained in terms of the CSV change in the structure  of
bound $p$'s and $n$'s.

\section{The Class IV NN Force}
In the terminology of Henley and Miller\cite{henley} a class IV force
affects only the $np$ system, mixing spin-singlet and triplet states.
Experimentally it is extremely difficult to detect such a mixing which
gives rise to a difference between the asymmetry measured in
$\vec{n} p$ and $\vec{p} n$ scattering (i.e. $\Delta A = A_n(\theta) -
A_p(\theta) \neq 0$) at the $10^{-3}$ level. The experimental determination of
this CSV at TRIUMF\cite{triumf} and IUCF\cite{iucf}
has been a superb achievement and some of this is captured in the
presentation of van Oers at this meeting\cite{vanoers}.

The theoretical contributions to the class IV interaction have been well
understood for some time\cite{miller,hht,gersten} -- at least within the
framework of one-boson-exchange forces. There is a characteristic
difference in the energy dependence of the contribution of the CSV force
arising from the $np$ mass difference at the nucleon vertex when a
charged, isovector meson is omitted (proportional to
($\underline{\tau}_1 \times \underline{\tau}_2)_z (\vec{\sigma}_1 \times
\vec{\sigma}_2) \cdot \vec{L}$) and that arising from $\gamma$-exchange
or $\rho-\omega$ mixing (proportional to ($\underline{\tau}_1
- \underline{\tau}_2)_z (\vec{\sigma}_1 -
\vec{\sigma}_2) \cdot \vec{L}$). At energies above $300 MeV$, where the
TRIUMF experiments have been performed, there is essentially no
sensitivity to the latter contribution, while at the energy of the IUCF
experiment $\Delta A$ is $20$ without, and $35$ with, $\rho-\omega$ mixing --
in comparison with the experimental value of $33 \pm 5.9 \pm 4.3$ (all
in units of $10^{-4}$).

Clearly the experiments are complementary, with
the second generation TRIUMF experiment confirming the prediction of
Holinde {\it et al.}\cite{hht} -- see also Ref.\cite{in} -- quite
precisely and thus confirming our understanding of the pion exchange
component of the NN force. The effect of $\rho-\omega$ mixing is then
confirmed by the IUCF measurement, but only at the level of $1-2\sigma$.
In view of the theoretical interest surrounding $\rho-\omega$ mixing,
which we describe next, and especially its vital role in the conventional
treatment of CSV in mirror nuclei\cite{mns,bi},
it is extremely important to make a
second generation experiment at an energy below $200 MeV$!

\subsection{Meson Mixing and Vector Meson Dominance}
The mixing between a ``real'' $\rho$ and $\omega$ is, of course,
observed in the measurement of the pion form-factor in $e^+-e^-$
annihilation. However, in calculating the usual CSV NN potential it is
assumed that there is no variation of the mixing amplitude from
$m_{\rho}^2$ to the space-like region (where it is needed to construct
the potential) {\it and} that there is no CSV at the NN$\rho$ or
NN$\omega$ vertices\cite{henley,mcn}.
Goldman {\it et al.} (GHT)\cite{GHT}
were the first to ask whether it was reasonable to assume that the
$\rho-\omega$ mixing amplitude is independent of $q^2$ and there has
since been a considerable body of work.

The initial GHT model was relatively simple. The vector mesons
were assumed to be quark-antiquark composites, and the mixing was
generated entirely by the small mass difference
between the up and down quark masses.
The mesons coupled to the quark loop via a form-factor
which modelled the meson substructure. Free
Dirac propagators were used for the quarks,
thus ignoring the question of confinement.
More recent work\cite{KTW,MTRC}
has modelled confinement by using quark propagators which are entire
(i.e. which do not have a pole in the finite complex-$q^2$ plane
so that the quarks are never on mass-shell).
The vector mesons couple to conserved currents which, as shown
by O'Connell {\it et al.}\cite{HOC},
leads to a node in the mixing amplitude when the momentum
($q^2$) of the meson vanishes.

The use of an intermediate nucleon loop\cite{PW} as the mechanism
driving $\rho-\omega$ mixing amplitude
(relying on the mass difference between the neutron and proton)
avoids the worries of quark confinement,
as well as enabling one to use well-known
parameters in the calculation (masses, couplings, etc).
This model has a node
for the mixing at $q^2=0$. Mitchell {\it et al.}\cite{MTRC} concluded
that in their bi-local theory
(where the meson fields are composites of quark
operators, e.g. $\omega_\mu(x,y)\sim \overline{q}(y)i\gamma_\mu q(x)$)
the quark loop mechanism alone generates an insignificant CSV
potential.

Iqbal and Niskanen\cite{IN} studied the effect of a $\rho-\omega$
mixing amplitude that vanished at $q^2 = 0$ on the CSV $np$ potential
and concluded (as GHT had suggested) that it reduced the effect to a
negligible level. The idea that the mixing amplitude should vanish at
$q^2 = 0$ was challenged on the grounds that for the same reasons the
$\gamma^*\rho$ coupling should vanish there and this would destroy the
phenomenological success of vector meson dominance (VMD)\cite{vanmil}.
In a recent
review of VMD, O'Connell {\it et al.}\cite{rev}
show that while this would be true
in the traditional form of VMD, in the original form introduced by
Sakurai\cite{sak} (VMD1 in the notation of O'Connell {\it et al.})
it is fact quite natural for the coupling to
vanish at $q^2 = 0$. In order to guarantee the equivalence of the two
formulations one must add a direct photon-hadron coupling in the VMD1
form. That VMD1 can also produce an excellent description of the
pion form-factor has recently been shown explicitly\cite{pionff}.
Indeed, it is only
in the older version of VMD that one can naturally include any deviation
from universality -- as observed in nature\cite{universality}.

Returning to the $\rho-\omega$ mixing potential we note that a
completely consistent calculation must deal with not only the mixing
amplitude but also with the vertices\cite{malt,tomjerry}
-- the independent parts of the
full calculation are dependent on the choice of interpolating fields for
the vector mesons. Gardner {\it et al.}\cite{gard} have recently
shown that for a very natural choice of interpolating field (the quark
vector current) there is a sizeable CSV at the vertices. This may
restore some of the CSV required. Their result is particularly important
because for their choice of vector meson fields the result of O'Connell
{\it et al.}\cite{HOC} shows that the mixing amplitude would indeed
vanish at $q^2 = 0$. In conclusion, we note that the one further
uncertainty over the CSV potential arising from $\rho-\omega$ mixing
is the range of the
form-factor used at the NN$\rho$ or NN$\omega$ vertex. If this is as
soft as suggested by the work of Deister {\it et al.}\cite{gari}, for
example, the $\rho-\omega$ mixing potential would still be negligible.
All of this simply increases the desperate need for a new measurement
below $200 MeV$.

\section{The Okamoto-Nolen-Schiffer Anomaly}
The Okamoto-Nolen-Schiffer (ONS) anomaly\cite{ok,ns} is a long-standing
problem in nuclear physics.  The anomaly is the discrepancy
between experiment and theory for the binding energy differences
of mirror nuclei -- after the removal of electromagnetic corrections.
Conventional nuclear contributions to the
anomaly are thought to be at the few per
cent level and cannot explain the experimental findings\cite{ns,shlomo}.
The effects of charge symmetry breaking in
the nuclear force\cite{ok,hs}, especially $\rho$-$\omega$ mixing,
seem to reproduce much of the discrepancy\cite{mns,bi,reex}, at least in light
nuclei. However, the investigations of
the off-shell variation of the $\rho$-$\omega$ mixing amplitude,
discussed in the previous section,
have put this explanation into question. As a consequence there has been
considerable interest in the development of alternative, quark-based,
approaches to the problem.

One of the earliest quark-based treatments of the ONS anomaly
was by Henley and Krein\cite{hk}. Using the
Nambu--Jona-Lasinio (NJL) model, they
indicated that the anomaly might be related to the partial
restoration of chiral symmetry in nuclear matter. More recent
theoretical investigations have involved QCD
sum-rules\cite{hb} and the quark cluster model\cite{nakamura}. In the
latter case, Nakamura {\it et al.} used a quark cluster model of the NN
force to incorporate the CSV effect of a quark mass difference in the
one-gluon-exchange hyperfine force in a study of nuclei in the $1s-0d$
shell. This is an ambitious program but the initial results look
promising.

We would like to briefly report on the application of the
quark-meson coupling (QMC) model of Guichon\cite{guichon} to this
problem. In this model, nuclear matter consists of non-overlapping
nucleon bags bound by the self-consistent exchange of
$\sigma$ and $\omega$ mesons in the mean-field approximation.
It has been extended to include the $\rho$ and
an isovector-scalar meson (the $\delta$)\cite{sath,ons}.
As well as providing an excellent description of the properties of
nuclear matter, it has been applied successfully to the calculation of
nuclear structure functions\cite{saito}. Furthermore, the relationship
between the QMC model and Quantum Hadrodynamics (QHD)\cite{serot} has been
investigated.  The fascinating result is that for infinite nuclear
matter the two approaches can be written in an identical form,
except for the appearance of the quark-scalar density in the self-consistency
condition for the scalar field\cite{sath}. The simplicity of
this finding suggests that it may be rather more general than the
specific model within which it was derived.

Retaining only the $\sigma$ mean field (which gives the dominant effect)
the main result of the model for the ONS anomaly is\cite{ons}:
\begin{equation}
\Delta_{np}^{\star} = \Delta_{np}^0 - g_{\sigma}
(C_n^{\sigma} - C_p^{\sigma}) \bar{\sigma}. \label{main}
\end{equation}
Here $\Delta_{np}^{\star}$ is the $n-p$ mass difference in matter of
density $\rho_B$, $\Delta_{np}^0$ is the free mass difference
and we have assumed symmetric nuclear matter. The dominant physics arises from
the nucleon internal structure which means that the $\sigma$ couples to
the nucleon through its scalar density $C_j^{\sigma}$ ($j=n,p$), which is
density dependent and larger for $j=n$ than for $j=p$ because of the
greater mass of the $d$-quark.
Because $C_n^{\sigma} > C_p^{\sigma}$
the {\it n-p\/} mass
difference decreases as $\rho_B$ goes up.

It is not possible to make an accurate calculation of the CSV effects
for finite nuclei using a model of infinite nuclear matter. Nevertheless
we can get a qualitative idea using local-density approximation. As seen
in Table\ref{nsaf} both the sign and the order of magnitude of the
anomaly are well reproduced.
\begin{table}[t]
\begin{center}
\caption{Estimate of the ONS anomaly (in MeV) for several finite nuclei using
local density approximation -- from Ref.$^{37}$. ($R_0$ is the bag
radius for the free nucleon.)}
\label{nsaf}
\begin{tabular}[t]{ccccc}
\hline
$R_0 (fm)$ & 0.6 & 0.8 & 1.0 & observed discrepancy \\
\hline
$^{15}$O--$^{15}$N & 0.29 & 0.32 & 0.34 & 0.16 $\pm$ 0.04 \\
$^{17}$F--$^{17}$O & 0.22 & 0.25 & 0.27 & 0.31 $\pm$ 0.06 \\
$^{39}$Ca--$^{39}$K & 0.36 & 0.41 & 0.44 & 0.22 $\pm$ 0.08 \\
$^{41}$Sc--$^{41}$Ca & 0.34 & 0.38 & 0.41 & 0.59 $\pm$ 0.10  \\
$^{120}$Sn & 0.72 & 0.83 & 0.87 &  \\
$^{208}$Pb & 0.78 & 0.91 & 0.95 & $\sim$ 0.9 \\
\hline
\end{tabular}
\end{center}
\end{table}

In conclusion, we note that although equ.\ref{main} was
derived within the QMC model it may be a more general result.
We see that the internal structure of the nucleon is crucial to the
understanding of the ONS anomaly in any relativistic model of nuclear
structure involving a scalar field. In particular, if
the quarks are relativistic and the {\it n-p\/} mass difference arises
because $m_u \neq m_d$, then in matter this mass difference will vary by
an amount proportional to $m_d - m_u$ and $\bar{\sigma}$. This variation
necessarily has the correct sign and magnitude to explain the ONS
anomaly. By comparison the magnitude of the CSV induced by the $n-p$
mass difference in QHD is an order of magnitude too small\cite{juelich}.

\section{Unitarity of the Cabibbo-Kobayashi-Maskawa Mass Matrix}
As we have heard at this meeting\cite{wein} it
is very important to refine our understanding of the weak
coupling to quarks. A violation of unitarity of the
Cabibbo-Kobayashi-Maskawa (CKM) matrix would be a clear indication of physics
beyond the standard model. The precision required for such a test,
particularly for the dominant matrix element, $V_{ud}$, presents a
tremendous challenge to experimenters and theorists alike.
In particular, the most accurate
experimental measurement of
the vector coupling constant in nuclear beta-decay comes from
super-allowed $0^+$-$0^+$ transitions between nuclear isotriplet states.
In order to relate these precise measurements to the
quark-level vector coupling, $V_{ud}$, one needs to apply a number of
small nuclear structure corrections\cite{rad} in addition to the relatively
standard radiative corrections\cite{town}. Despite intensive study of
these nuclear ``mismatch'' corrections\cite{mf1,mf2} there
remains a systematic difference of a few tenths of a percent
between the value of $V_{ud}$ inferred
from the vector coupling measured in muon decay, $G_{\mu}$, and
unitarity of the CKM matrix and those determined from the nuclear
$ft$-values. For recent summaries we refer to the reviews of
Wilkinson\cite{wilk} and Towner and Hardy\cite{towner}, and also
to the recent report by Savard {\it et al.}\cite{savard} of accurate
data on $^{10}$C.

Until now the nuclear corrections have been explored within the
framework of conventional nuclear theory with point-like nucleons.
Of course, for the nucleon itself there has been considerable
investigation of the effect on the vector form-factor of the breaking of
CVC caused by the small $u$-$d$ mass difference in QCD\cite{ad,bs}.
While this is necessarily very small, the measurements of $V_{ud}$ and
$G_{\mu}$ are also extremely precise. Thus we have been led to ask
whether this small nuclear discrepancy might be associated with a change
in the degree of non-conservation of the vector current
caused by nuclear binding\cite{ckm}.

In order to investigate whether nuclear binding might influence the Fermi
decay constant of the nucleon itself one needs a model of nuclear
structure involving explicit quark degrees of freedom which nevertheless
provides an acceptable description of nuclear binding and saturation.
The QMC model, described in the previous section,
seems ideally suited to the problem. It allows us to examine the
variation with density of the quark vector current
matrix element:
\begin{equation}
I_{ii'}(\rho_B) = \int_{Bag} dV \psi_{i/p}^{\dagger} \psi_{i'/n},
\label{Iji}
\end{equation}
with $i'=d$ and $i=u$ for the $d \rightarrow u$ conversion and $i=i'=u$ or
$d$ for the two spectator quarks. As the radius of the proton and
neutron are different we integrate over the common volume.

The decrease in $I_{ii'}$ as
the density increases is a direct consequence of the increasing
difference between the proton and neutron radii -- that is the smaller
volume of overlap. In the calculation of Saito and Thomas\cite{ckm}
the deviation of
$I_{ii'}(\rho_B)/I_{ii'}(0)$ from unity is roughly linear with density:
\begin{equation}
\frac{I_{ii'}(\rho_B)}{I_{ii'}(0)} \simeq 1 - a_{ii'} \times
\left( \frac{\rho_B}{\rho_0}
\right), \label{Iii}
\end{equation}
with $a_{ii'} \simeq (2.4, 2.9, 3.3) \times 10^{-4}$ for $R_0$ =
(0.6, 0.8, 1.0) fm, respectively, (for any combination of $ii'$) and
$\rho_0$ the normal nuclear density (0.17 $fm^{-3}$).

The evaluation of $ft$-values involves the inverse of the product of
$I_{ud}, I_{uu}$ and $I_{dd}$ squared.  Since for a given,
free (average) radius
of the bag each of these matrix elements
decreases by roughly the same amount, the fractional
increase in the $ft$-value with
density is therefore
\begin{equation}
\frac{ft(\rho_B)}{ft(0)} \simeq 1 + b \times
\left( \frac{\rho_B}{\rho_0}
\right), \label{ft}
\end{equation}
with $b$ approximately six times the decrease in each
integral -- i.e. $b \simeq (1.5, 1.8, 2.0) \times
10^{-3}$ for $R_0$ = (0.6, 0.8, 1.0) fm.
Thus the {\it increase} in the $ft$-value at $\rho_0/2$ ranges from $0.075\%$
to $0.10\%$, while at $\rho_0$ it lies between $0.15\%$ and $0.20\%$.
This is to be compared with a violation of unitarity of the CKM matrix
of $0.35 \pm 0.15\%$ in the most recent
analysis of Towner and Hardy\cite{towner}.

While it is not possible to
draw unambiguous conclusions from a comparison of
theoretical results in infinite nuclear matter with data from finite
nuclei, these results are extremely encouraging.
At $\rho_0/2$ the calculation suggests a reduction in the
violation of unitarity by about $1/3$, while at $\rho_0$ a correction as
big as $0.2\%$ brings the discrepancy back to only one standard
deviation.

The essential physics involved in this calculation is CSV,
in particular, the fact that {\it in nuclear matter}
the confining potential felt by a
quark in a proton is not the same as that felt by a quark
in a neutron.
We have already  explained that a relativistic field theory only yields the
right order of magnitude for nuclear charge symmetry breaking if the
relevant mass scale involves quarks rather than nucleons\cite{juelich}.
In this sense the ONS anomaly may prove to
be something of a ``smoking gun'' for
quark degrees of freedom in nuclei. This is even more obvious here; it
is only because the nuclear charge symmetry violation occurs at the
quark level that it can produce a deviation of the vector form factor of
the bound nucleon from its free value.

\section{Conclusion}
In this brief review we have seen that charge symmetry violation
provides a very specific and powerful tool to probe the nuclear force.
Through studies in the NN system we have been led to a deeper
understanding of $\rho-\omega$ mixing and indeed of vector dominance
itself. In struggling to understand the ONS anomaly in mirror nuclei we
have confronted the role of quark degrees of freedom in nuclei. As we
have seen, a treatment of nuclear structure at the quark level may also
be required to understand the apparent violation of unitarity for the
Cabibbo-Kobayashi-Maskawa matrix when $V_{ud}$ is extracted from
super-allowed Fermi beta-decay. There can be no doubt that further study
of charge symmetry violation in hadronic systems will continue to
provide a wealth of information on strong interaction dynamics.

\section{Acknowledgements}
This work was supported in part by the Australian Research Council.

\end{document}